\documentclass[twocolumn]{aastex62}

\usepackage{amsmath}
\usepackage{multirow}
\usepackage{hyperref}
\usepackage{bm}

\shorttitle{Analysis of the duration--hardness ratio plane of gamma-ray bursts using skewed distributions}
\shortauthors{Mariusz Tarnopolski}

\begin{document}

\title{Analysis of the duration--hardness ratio plane of gamma-ray bursts using skewed distributions}

\author{Mariusz Tarnopolski}
\email{mariusz.tarnopolski@uj.edu.pl}

\affiliation{Astronomical Observatory, Jagiellonian University, Orla 171, 30--244, Krak\'ow, Poland}

\begin{abstract}

The two widely accepted classes of gamma-ray bursts (GRBs), short and long, are with confidence ascribed to mergers of compact objects and collapse of massive stars, respectively. A third, intermediate/soft class, remains putative. Its existence was claimed based on univariate and bivariate analyses of GRB observables modeled with Gaussian distributions. This, however, may not be the appropriate approach, as it has been already shown that the univariate distributions of durations are better described by mixtures of two skewed components rather than three Gaussian ones.

This paper investigates whether data in the duration--hardness ratio plane is better modeled by mixtures of skewed bivariate distributions than by normal ones. The archival data set of the {\it Compton Gamma-Ray Observatory}/BATSE and {\it Fermi}/GBM data from the most recent catalogue release are examined. The preferred model is chosen based on two information criteria, Akaike ($AIC$) and Bayesian ($BIC$). It is found that the best description is given by a two-component mixture of skewed Student-$t$ distributions, which outperforms any other model considered. This implies that the distribution of the studied parameters is intrinsically skewed, introducing spurious Gaussian components, and hence the third class is unlikely to be a real phenomenon. Its existence, based on statistical inference, is therefore rejected as unnecessary to explain the observations.

\end{abstract}


\section{Introduction}
\label{sect1}

Gamma-ray bursts \citep[GRBs,][]{klebesadel} were early recognized to have a bimodal duration distribution \citep{mazets}. The division between short \citep{eichler89,paczynski91,narayan92} and long GRBs \citep{woosley93,paczynski98,macfadyen99} was established, based on the distribution of $T_{90}$ (the time during which 90\% of the GRB's fluence is accumulated, starting from the time at which 5\% of the total fluence is detected), to be at $T_{90} \simeq 2\,{\rm s}$ \citep[\citealt{kouve93}; but see also][]{fynbo,king,kann,bromberg,tarnopolski15a,tarnopolski15c,li2}. The progenitors of short GRBs are believed to be double neutron star (NS-NS) or NS-black hole (BH) mergers \citep{nakar}. The association of a kilonova with GRB 130603B provided strong evidence for the nature of the progenitors of short GRBs \citep{tanvir13}. The recent detection of the gravitational wave event together with a short GRB as its electromagnetic counterpart, GW/GRB170817, further confirmed the relation between short GRBs and compact-object mergers \citep{abbott1,abbott2,goldstein,savchenko}. The progenitors of long GRBs are associated with supernovae Ic \citep{filippenko} related with collapse of massive, e.g. Wolf-Rayet or blue supergiant, stars \citep{galama,hjorth,stanek,woosley06,cano,perna}. No connection between short GRBs and supernovae has been proven \citep{zhang09,ruffini}. 

The durations $T_{90}$ were early noticed \citep{mcbreen,koshut,kouve96} to roughly follow a log-normal distribution (i.e., $\log T_{90}$ to be normal), and routinely fitted thereafter as such. \citet{horvath98} found a prominent third peak, between the short and long groups, in the $\log T_{90}$ distribution of GRBs detected by the Burst And Transient Explorer onboard the {\it Compton Gamma-Ray Observatory} \citep[{\it CGRO}/BATSE;][]{meegan92,paciesas}, and hence claimed the existence of an intermediate-duration class of GRBs. However, when more data was accumulated, this peak blended into the bulk of the distribution, manifesting itself only as a small bump on the shorter side of the long GRBs group \citep{horvath02,tarnopolski15b}, adding to the skewness of the component. The evidence for a third normal component in $\log T_{90}$ was found also in {\it Swift} Burst Alert Telescope (BAT) data \citep{horvath08,horvath10,zhang08,huja,zitouni,zhang16}. {\it Swift} GRBs form the largest sample of GRBs with measured redshifts observed by the same instrument, making the analysis in both observer and rest frames possible. It was found \citep{huja,tarnopolski16b,zhang16,kulkarni} that three and two Gaussian components are required in the observer and rest frames, respectively; however, \citet{zitouni} found three groups in both frames. Interestingly, only two components are required for the BATSE dataset in the observer frame \citep{zitouni,zhang16}, contrary to the findings of \citet{horvath02}. \citet{kulkarni} also did not find decisive evidence for a third component in case of BATSE. Regarding {\it Fermi} Gamma-ray Burst Monitor (GBM, \citealt{kienlin,gruber,bhat}), \citet{bystrycky,bhat,zhang16,kulkarni} found that two components suffice for the logarithmic duration distribution to be adequately described. Using pseudo-redshifts derived from the $L_p-E_p$ relation \citep{yonetoku10,tsutsui13}, the same conclusion was reached by \citet{zitouni18}. Evidence for a third normal component was found, however, in the {\it RHESSI} dataset \citep{ripa09}. In the data from {\it BeppoSAX} \citep{frontera}, due to low sensitivity to short GRBs (caused by a 1-second short integration time), there were only two peaks in the $\log T_{90}$ distribution, corresponding to intermediate and long classes \citep{horvath09}. In case of {\it Suzaku} Wide-band All-sky Monitor (WAM), a two-component mixture of lognormal distributions is favored over a three-component one \citep{ohmori}.

It was argued \citep{koen,tarnopolski15b} that the logarithmic duration distribution need not necessarily be normal; the assymmetry (skewness) can originate from, e.g., an asymmetric distribution of the progenitor envelope mass \citep{zitouni}. Therefore, the BATSE, {\it Swift} and {\it Fermi} data sets were examined previously with skewed distributions \citep{tarnopolski16a,tarnopolski16c,kwong}. The reasoning is that modeling an inherently skewed distribution with a mixture of symmetric ones requires excessive components to be included, resulting in a spurious determination of the number of underlying classes \citep{koen}. It is conceptually and technically easier to introduce an additional parameter in the modeling of short and long GRBs rather than to invent a new physical mechanism giving rise to an elusive intermediate class. It was indeed found that mixtures of two skewed components are either significantly better than, or at least as good as, three-component symmetric models, meaning that the third class is discarded as unnecessary \citep{tarnopolski16a}. Moreover, a careful analysis of the properties of the presumed class of intermediate GRBs showed that they differ from long GRBs only in having lower luminosities \citep{ugarte}, so that they might be simply a low-luminosity tail of the long GRBs group.

A univariate analysis cannot, though, reveal all the intricacies of separating GRBs into meaningful classes. A natural step is to examine a two-dimensional realm of the $T_{90}-H$ plane composed of the duration and ratio of fluences in two energy bands (i.e., hardness ratio). \citet{mukh,horvath06} (with BATSE data), \citet{ripa09} ({\it RHESSI}), \citet{horvath10,veres} ({\it Swift}) performed analyses of the $T_{90}-H$ distribution similarly to the univariate case, i.e. assumed a bivariate Gaussian mixture model and seeked the number of components that fits the data best; they all found a three-component model to be more favorable than that with two components. \citet{ripa12}, however, arrived at only two components in case of the {\it RHESSI} data set. \citet{horvath12} performed a principal component analysis that was followed by fitting mixtures of bivariate Gaussian distribution; it was found that a three-component model is the optimal one in terms of goodness of fit. On the other hand, \citet{yang} examined {\it Swift} GRBs with measured redshift, and showed that two components suffice in both the observer and rest frames. For the {\it Fermi} sample, contradictory results have been obtained: \citet{bhat} arrived at two, while \citet{horvath18} at three components as the most favorable.

Several classifications were done in higher-dimensional parameter spaces. \citet{mukh} performed nonparametric and multinormal clustering of 797 BATSE 3B GRBs in a space of 6 parameters (durations $T_{90}$ and $T_{50}$, defined in a similar fashion as $T_{90}$, peak flux measured in 256 ms bins $P_{256}$, total fluence $F_{\rm tot}$, and hardness ratios $H_{32}$ and $H_{321}$). The nonparametric approach yielded ambiguous results, pointing at two or three clusters, while the multinormal modeling (in a three-dimensional space of $T_{90}$, total fluence and $H_{321}$---the hardness ratio $H_{32}$ was eliminated) indicated the GRB population consisted of three classes. It should be emphasized that the examined data set was the same as in \citep{horvath98}, where a prominent third peak was discovered in the duration distribution, but dissapeared when more data was accumulated. \citet{balastegui} claimed the existence of a third class based on neural network classification. However, \citet{hakkila00,hakkila03} attributed the presence of this class to instrumental effects, and questioned its physical reality; this conclusion was also supported by \citet{rajaniemi}, who employed an independent analysis method (self-organizing map, \citealt{kohonen}). The outputs of such unsupervised classifications are affected by several factors, e.g. the employed technique, specific samples and attributes used, among others \citep{hakkila04}, and by systematic biases \citep{roiger00}. \citet{chattopadhyay07} on the other hand used different clustering methods ($K$-means and Dirichlet process; the latter with an underlying assumption of a multinormal distribution), and again found statistical evidence for three GRB classes. \citet{veres} claimed, based on the $K$-means method as well, to find evidence for the third class, too. The same approach turned out to be inconclusive for the {\it RHESSI} data \citep{ripa12}; on the other hand, multinormal fitting in the three-dimensional space of $T_{90}$, $H$ and peak-count rates yielded three components.

\citet{chattopadhyay17} examined the complete BATSE data in a six-dimensional space of the same parameters as \citet{mukh}. By means of a multivariate Gaussian mixture model, they arrived at a conclusion that there are five clusters in this space. The same result was achieved by modeling with a multivariate Student-$t$ distribution \citep[][but see also Sect.~\ref{sect3.1} herein]{chattopadhyay18}. \citet{acuner} employed the Gaussian mixture model to analyze {\it Fermi} GRBs in a different space of the \citet{band} spectral parameters $(\alpha,\beta,E_{\rm peak})$, the duration $T_{90}$ and the fluence, and also claimed evidence for five groups.

While not of direct importance herein, it is worth to mention that the GRB family, besides short and long bursts, includes also ultra-long GRBs \citep{gendre,levan,zhang14,perna}, low-luminosity GRBs \citep{bromberg11}, and short GRBs with extended emission \citep{norris06,kaneko15}, i.e. having durations that would classify them as long GRBs, but without an associated supernova. They most likely originate from the merger of a white dwarf with an NS \citep{king} or BH \citep{dong}.

The aim of this work is to analyze two most numerous GRB samples, the {\it CGRO}/BATSE and {\it Fermi}/GBM data sets, in the two-dimensional space of $T_{90}-H$, using mixtures of skewed distributions, in order to establish the number of GRB classes. This is the first, except for \citep{tarnopolski16a}, attempt to model GRB groups with skewed distributions\footnote{Skewed distributions have been, however, employed in other astrophysical applications, e.g. in modeling the mass distribution of neutron stars \citep{kiziltan13}.}, and first such approach in a bivariate scheme. In Sect.~\ref{sect2} the examined data sets and statistical methods together with the employed probability distributions are briefly described. Sect.~\ref{sect3} presents the results obtained for both GRB samples. Sect.~\ref{sect4} is devoted to discussion, and concluding remarks are gathered in Sect.~\ref{sect5}. The \texttt{R} software\footnote{\url{http://www.R-project.org/}} is utilized throughout; the fittings are performed with the package \texttt{mixsmsn}\footnote{\url{https://cran.r-project.org/web/packages/mixsmsn/index.html}} \citep{prates}.

\section{Datasets and methods}
\label{sect2}

\subsection{Samples}
\label{sect2.1}

The {\it Fermi} data set \citep{bhat} contains 1376 GRBs with measured both $T_{90}$ and $H_{32}$ (P. Veres, priv. comm.), where the hardness ratio $H_{32}=\frac{F_{50-300\,{\rm keV}}}{F_{10-50\,{\rm keV}}}$ is the ratio of fluences in the respective energy bands during the $T_{90}$ interval. {\it CGRO}/BATSE\footnote{\url{https://gammaray.nsstc.nasa.gov/batse/grb/catalog/current/}} contains 1954 GRBs with $T_{90}$ and $H_{32}$, where the hardness ratio is computed with slightly different energy bands: $H_{32}=\frac{F_{100-300\,{\rm keV}}}{F_{50-100\,{\rm keV}}}$.

\subsection{Statistical methods}
\label{sect2.2}

\subsubsection{Maximum loglikelihood fitting}
\label{sect2.2.1}

Having a distribution\footnote{Bivariate distributions are considered herein, but the methodology is applicable for any dimensionality of the data; see \citep{tarnopolski16a,tarnopolski16b,tarnopolski16c} for a univariate analysis of $T_{90}$.} with a probability density function (PDF) given by $f=f(\bm{x};\theta)$ (possibly a mixture), where $\theta=\left\{\theta_i\right\}_{i=1}^p$ is a set of $p$ parameters, the log-likelihood function is defined as
\begin{equation}
\mathcal{L}_p(\theta)=\sum\limits_{i=1}^N\ln f(x_i;\theta),
\label{eq1}
\end{equation}
where $\left\{\bm{x}_i\right\}_{i=1}^N$ are the datapoints from the sample to which a distribution is fitted. The fitting is performed by searching a set of parameters $\hat{\theta}$ for which the log-likelihood is maximized \citep{kendall}. When nested models are considered, the maximal value of the log-likelihood function $\mathcal{L}_{p,\rm max}\equiv\mathcal{L}_p(\hat{\theta})$ increases when the number of parameters $p$ increases.

\subsubsection{Model comparison---information criteria}
\label{sect2.2.2}

For nested as well as non-nested models, the information criteria (IC): Akaike IC ($AIC$) and Bayesian IC ($BIC$) may be applied \citep{akaike,schwarz,burnham,biesiada,liddle,tarnopolski16a,tarnopolski16b}. They are defined as
\begin{equation}
AIC=2p-2\mathcal{L}_{p,\rm max}
\label{eq2}
\end{equation}
and
\begin{equation}
BIC=p\ln N-2\mathcal{L}_{p,{\rm max}}.
\label{eq3}
\end{equation}
A preferred model is the one that minimizes $AIC$ or $BIC$. The expressions for both IC consist of two competing terms: the first measuring the model complexity (number of free parameters) and the second measuring the goodness of fit (or more precisely, the lack of thereof). The formulation of these IC penalizes the use of an excessive number of parameters. It prefers models with fewer parameters, as long as the others do not provide a substantially better fit. In case of $BIC$, the penalization term is greater than the corresponding term from the $AIC$, $p\ln N>2p$, for $N\geq 8$. Hence, the penalization in case of the $BIC$ is much more stringent, especially for large samples. 

What is essential in assesing the goodness of a fit in the $AIC$ method is the difference, $\Delta_i=AIC_i-AIC_{\rm min}$. If $\Delta_i<2$, then there is substantial support for the $i$-th model (or the evidence against it is worth only a bare mention), and the proposition that it is a proper description is highly probable. If $2<\Delta_i<4$, then there is strong support for the $i$-th model. When $4<\Delta_i<7$, there is considerably less support, and models with $\Delta_i>10$ have essentially no support \citep{burnham,biesiada}. It is important to note that when two models with similar $\mathcal{L}_{\rm max}$ are considered, the $\Delta_i$ depends solely on the number of parameters due to the $2p$ term in Eq.~(\ref{eq2}). Hence, when $\Delta_i/(2\Delta p)<1$, the relative improvement is due to actual improvement of the fit, not to increasing the number of parameters only.

In case of $BIC$, $\Delta_i=BIC_i-BIC_{\rm min}$, and the support for the $i$-th model (or evidence against it) also depends on the differences: if $\Delta_i<2$, then there is substantial support for the $i$-th model. When $2<\Delta_i<6$, then there is positive evidence against the $i$-th model. If $6<\Delta_i<10$, the evidence is strong, and models with $\Delta_i>10$ yield a very strong evidence against the $i$-th model \citep[essentially no support;][]{kass}.

Despite apparent similarities between the $AIC$ and $BIC$, it ought to be stressed that they answer different questions, as they are derived based on different assumptions. $AIC$ tries to select a model that most {\it adequately} describes reality (in form of the data under examination). This means that in fact the model being a real description of the data is never considered. On the contrary, $BIC$ tries to find the {\it true} model among the set of candidates. Because $BIC$ is more stringent, it has a tendency to underfit (resulting in an excessively simple model), while $AIC$, as a more liberal method, is inclined towards overfitting (accepting more parameters than needed). This may lead to pointing different models by the two criteria, which happens rarely, but is due to the fact that they try to satisfy different conditions.

\subsection{Distributions}
\label{sect2.3}

A mixture of $n$ components, each having a PDF given by $f_i(\bm{x};\theta^{(i)})$, is defined as
$$f(\bm{x};\theta) = \sum\limits_{i=1}^n A_i f_i(\bm{x};\theta^{(i)})$$
with the weights satisfying $\sum_{i=1}^n A_i = 1$, and $\theta = \bigcup_{i=1}^n \theta^{(i)}$. The following distributions are considered.

The multivariate, $k$-dimensional normal (Gaussian) distribution has a PDF:
\begin{equation}
\varphi_k(\bm{x};\bm{\mu},\bm{\Sigma}) = \frac{1}{\sqrt{(2\pi)^k|\bm{\Sigma}|}} \exp\left[ -\frac{1}{2}(\bm{x}-\bm{\mu})^\top \bm{\Sigma}^{-1} (\bm{x}-\bm{\mu}) \right],
\label{eq4}
\end{equation}
where $\bm{\mu}$ is the location vector (which in this case is also the mean, since the distribution is not skewed), $\bm{\Sigma}$ is the covariance matrix, and $|\bm{\Sigma}|=\det \bm{\Sigma}$. In particular, for a bivariate case ($k=2$),
\begin{equation}
\bm{\Sigma} = \left( \begin{array}{cc} \sigma_x^2 & \rho\sigma_x\sigma_y \\ \rho\sigma_x\sigma_y & \sigma_y^2 \end{array} \right).
\label{eq5}
\end{equation}
A mixture of $n$ components is described by $p=6n-1$ free parameters.

The multivariate skew-normal ($\mathcal{SN}$) distribution \citep{azzalini99,kollo,prates} is given by
\begin{equation}
f_k^{(\mathcal{SN})}(\bm{x};\bm{\mu},\bm{\Sigma},\bm{\lambda}) = 2 \varphi_k(\bm{x};\bm{\mu},\bm{\Sigma}) \Phi\left( \bm{\lambda}^\top\bm{\Sigma}^{-1/2}(\bm{x}-\bm{\mu}) \right),
\label{eq6}
\end{equation}
where $\Phi(.)$ denotes the CDF of a univariate standard normal distribution, and $\bm{\lambda}$ denotes the skewness parameter vector. If $\bm{\lambda}=\bm{0}$, then Eq.~(\ref{eq6}) reduces to Eq.~(\ref{eq4}). The mean of the $\mathcal{SN}$ distribution is $\bm{m}=\bm{\mu}+\sqrt{\frac{2}{\pi}}\frac{\bm{\Sigma}\bm{\lambda}}{\sqrt{1+\bm{\lambda^\top\bm{\Sigma}\bm{\lambda}}}}$, i.e. the location parameter $\bm{\mu}$ is not the mean itself, and the covariance is given by $\bm{\Sigma}-(\bm{m}-\bm{\mu})(\bm{m}-\bm{\mu})^\top$. The skewness\footnote{Multivariate measures of skewness are not as unambiguous as in the univariate case \citep{balakrishnan}, hence no explicit formulae are given herein.} is nonzero unless $\bm{\lambda}=0$. A mixture of $n$ components is described by $p=8n-1$ free parameters.

The multivariate Student $t$ ($\mathcal{T}$) distribution \citep{basso,cabral,prates} with $\nu$ degrees of freedom (dof) is defined to be
\begin{equation}
\begin{split}
f_k^{(\mathcal{T})}(\bm{x};\bm{\mu},\bm{\Sigma},\nu) &= \frac{1}{\sqrt{(\pi\nu)^k|\bm{\Sigma}|}} \frac{\Gamma\left(\frac{\nu+k}{2}\right)}{\Gamma\left(\frac{\nu}{2}\right)} \\
&\times \left( 1+\frac{1}{\nu}(\bm{x}-\bm{\mu})^\top \bm{\Sigma}^{-1} (\bm{x}-\bm{\mu}) \right)^{-\frac{\nu+k}{2}},
\end{split}
\label{eq7}
\end{equation}
where $\Gamma$ is the gamma function. The mean (for $\nu>1$) of the $\mathcal{T}$ distribution is $\bm{\mu}$, and the covariance matrix (for $\nu>2$) is $\frac{\nu}{\nu-2}\bm{\Sigma}$. In the limit $\nu\rightarrow\infty$, the $\mathcal{T}$ distribution approaches the normal distribution from Eq.~(\ref{eq4}). A mixture of $n$ components is described by $p=6n$ free parameters.

The multivariate skew-$\mathcal{T}$ ($\mathcal{ST}$) distribution \citep{kollo,cabral,prates} is defined as 
\begin{equation}
\begin{split}
&f_k^{(\mathcal{ST})}(\bm{x};\bm{\mu},\bm{\Sigma},\nu,\bm{\lambda}) = 2 f_k^{(\mathcal{T})}(\bm{x};\bm{\mu},\bm{\Sigma},\nu) \\
&\times T_{\nu+k}\left( \sqrt{\frac{\nu+k}{\nu+(\bm{x}-\bm{\mu})^\top \bm{\Sigma}^{-1} (\bm{x}-\bm{\mu})}} \bm{\lambda}^\top \bm{\Sigma}^{-1/2} (\bm{x}-\bm{\mu}) \right),
\end{split}
\label{eq8}
\end{equation}
where $T_{\nu+k}$ denotes the CDF of the standard univariate Student-$t$ distribution with $(\nu+k)$ dof, and $\bm{\lambda}$ is the skewness parameter vector. Eq~(\ref{eq8}) reduces to Eq.~(\ref{eq7}) for $\bm{\lambda}=0$. In the limit $\nu\rightarrow\infty$, the $\mathcal{ST}$ distribution approaches the $\mathcal{SN}$ distribution from Eq.~(\ref{eq6}). The mean (for $\nu>1$) of the $\mathcal{ST}$ distribution is $\bm{m} = \bm{\mu}+\bm{\omega}\bm{\xi}$, and its covariance (for $\nu>2$) is $\frac{\nu}{\nu-2}\bm{\Sigma}-(\bm{m}-\bm{\mu})(\bm{m}-\bm{\mu})^\top$, where $\bm{\xi}=\sqrt{\frac{\nu}{\pi(1+\bm{\lambda}^\top\bm{\Sigma}\bm{\lambda})}}\frac{\Gamma\left( \frac{\nu-1}{2} \right)}{\Gamma\left( \frac{\nu}{2} \right)} \bm{\Sigma}\bm{\lambda}$ and $\bm{\omega} = {\rm diag}(\Sigma_{11},\ldots,\Sigma_{kk})^{1/2} $ \citep{azzalini03,kollo}. The skewness (for $\nu>3$) is nonzero unless $\bm{\lambda}=0$. A mixture of $n$ components is described by $p=8n$ free parameters.

The distributions are referred to as: 2G and 3G for the mixture of two and three Gaussian components, respectively; 2$\mathcal{SN}$ and 3$\mathcal{SN}$ for the respective mixtures of $\mathcal{SN}$ distributions; 2$\mathcal{T}$ and 3$\mathcal{T}$ in case of the $\mathcal{T}$ distribution; and 2$\mathcal{ST}$ and 3$\mathcal{ST}$ for the $\mathcal{ST}$ distribution.

\section{Results}
\label{sect3}

\subsection{BATSE}
\label{sect3.1}

\begin{figure*}
\plotone{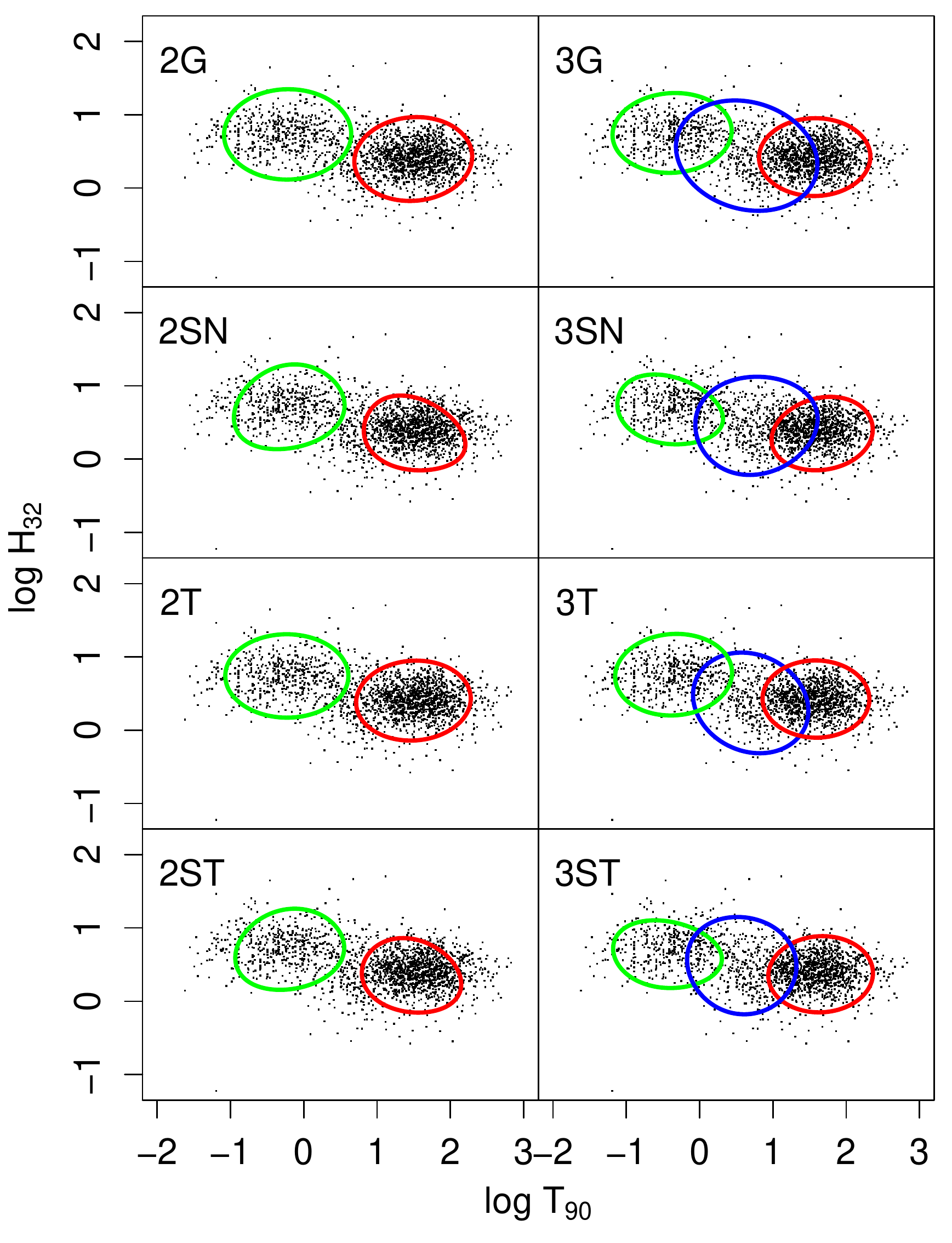}
\caption{Fits to the BATSE data. The contours depict the FWHM of each component, and $T_{90}$ is measured in seconds.}
\label{fig1}
\end{figure*}

\begin{figure}
\plotone{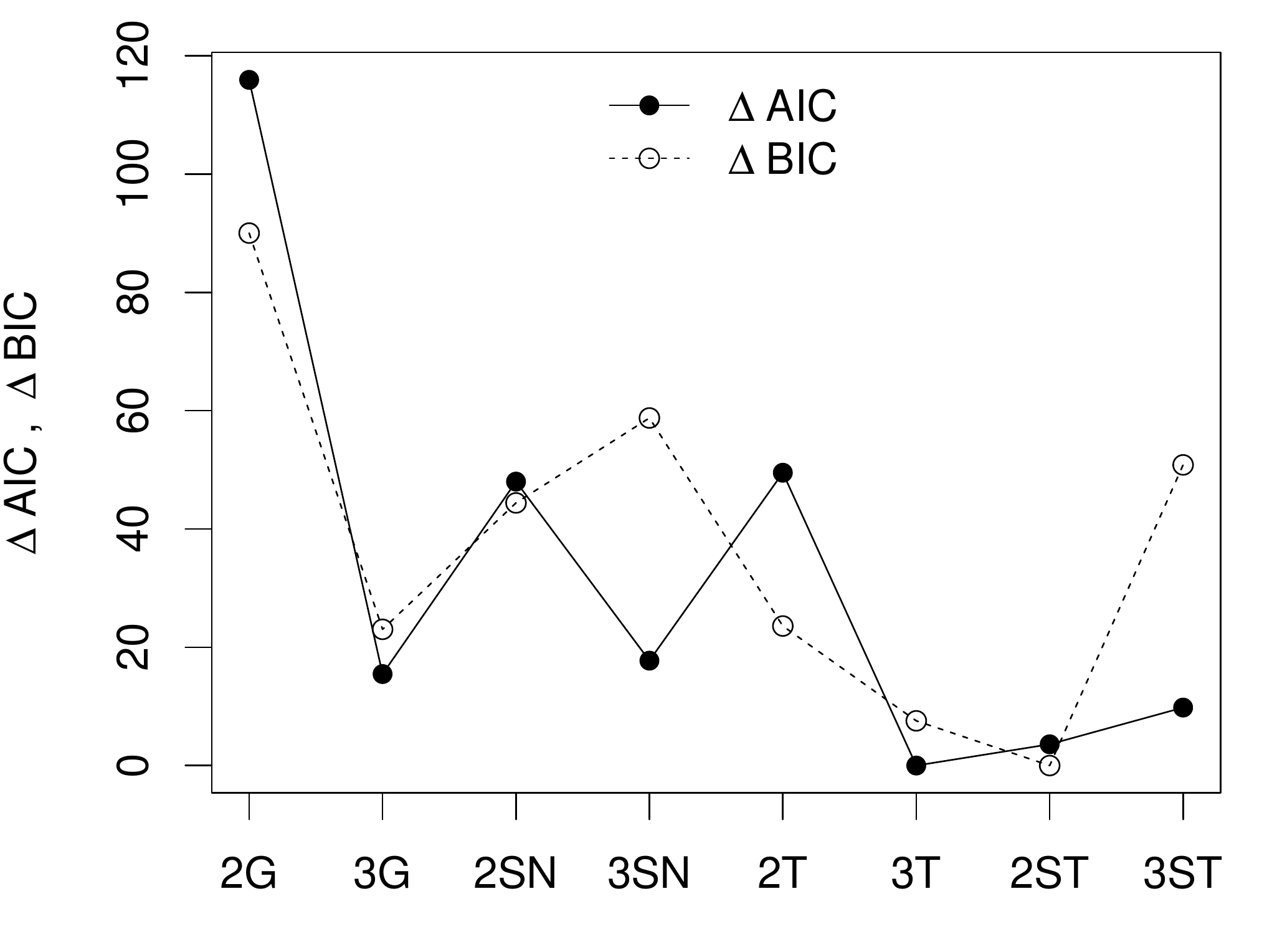}
\caption{Information criteria scores for the BATSE data.}
\label{fig3}
\end{figure}

Results of the fittings performed with the set of 1954 BATSE GRBs are displayed in graphical form in Fig.~\ref{fig1}, whereas the obtained parameters are gathered in Table~\ref{tbl1}, which contains also the values of $\mathcal{L}_{\rm max}$, $\Delta AIC$ and $\Delta BIC$. The BATSE data set consists in $\sim 25\%$ of short GRBs, forming a distinct from the long GRBs cloud in the $T_{90}-H_{32}$ plane, making the two-component fits consistent with each other qualitatively, as follows from the left column of Fig.~\ref{fig1}. Likewise, for all statistical models considered, introduction of a third component placed it roughly between the two major classes, without significantly affecting them. The $\Delta AIC$ and $\Delta BIC$ values, sorted in increasing order, are additionally gathered in Table~\ref{tbl2} for clarity, with a graphical representation in Fig.~\ref{fig3}. It follows that the $AIC$ points at the 3$\mathcal{T}$ model as the one that best describes the data, with the 2$\mathcal{ST}$ in the second position. Due to the $\Delta AIC = 3.6$, the support for the latter is strong. On the other hand, the $BIC$ points at 2$\mathcal{ST}$ as the best model, with the 3$\mathcal{T}$ with weak support ($\Delta BIC = 7.56$). Given that the $AIC$ ($BIC$) has a tendency towards overfitting (underfitting), and that the goal herein is to obtain the simplest model possible that adequately describes the data, overall the 2$\mathcal{ST}$ model is more likely to underly the observations.

Within the $AIC$ framework, the 3$\mathcal{ST}$ model (the most complex among the examined, with $p=24$ free parameters) is barely worth mentioning ($\Delta AIC = 9.8$), with the remaining models---in the $BIC$ framework as well---confidently rejected. In particular, the 2G is the worst model in both schemes, and the celebrated 3G is characterized with both $\Delta AIC$ and $\Delta BIC$ well above the value of 10. It is curious that the mixtures of the simplest skewed model---the $\mathcal{SN}$---perform rather poorly \citep[compare with][]{tarnopolski16a}.

\begin{table*}
\caption{Parameters of the fits to the BATSE data.}
\label{tbl1}
\centering
\begin{tabular}{c c c c c c c c c c}
\hline

Model & $A$ & $\bm{\mu}$ & $\bm{\Sigma}$ & $\bm{\lambda}$ & $\nu$ & $\mathcal{L}_{\rm max}$ & $\Delta AIC$ & $\Delta BIC$ & $p$ \\

  \hline

\multirow{3}{*}{2G} & 0.283 & $(-0.220, 0.732)$ & $\left( \begin{array}{cc} 0.543 & 0.005 \\ 0.005 & 0.274 \end{array} \right)$ & --- & \multirow{3}{*}{---} & \multirow{3}{*}{$-2415.561$} & \multirow{3}{*}{115.918} & \multirow{3}{*}{90.011} & \multirow{3}{*}{11} \\
                    & 0.717 & $(1.494, 0.397 )$ & $\left( \begin{array}{cc} 0.466 & 0.019   \\ 0.019 & 0.236   \end{array} \right)$ & --- &  &  &  &  &  \\

  \hline

   & 0.198 & $(-0.374, 0.752)$ & $\left( \begin{array}{cc} 0.476 & 0.017  \\ 0.017 & 0.214  \end{array} \right)$ & --- &     &  &  &  &  \\
3G & 0.189 & $(0.643, 0.442 )$ & $\left( \begin{array}{cc} 0.671 & -0.087 \\ -0.087 & 0.409 \end{array} \right)$ & --- & --- & $-2359.336$ & 15.469 & 23.028 & 17 \\
   & 0.613 & $(1.570, 0.423 )$ & $\left( \begin{array}{cc} 0.416 & 0.004  \\ 0.004 & 0.203  \end{array} \right)$ & --- &     &  &  &  &  \\

  \hline

\multirow{3}{*}{2SN} & 0.301 & $(-0.734, 0.836)$ & $\left( \begin{array}{cc} 0.857 & -0.069  \\ -0.069 & 0.272  \end{array} \right)$ & $(1.923, -0.651)$ & \multirow{3}{*}{---} & \multirow{3}{*}{$-2377.601$} & \multirow{3}{*}{47.998} & \multirow{3}{*}{44.402} & \multirow{3}{*}{15} \\

                     & 0.699 & $(1.866, 0.584)$  & $\left( \begin{array}{cc} 0.585 & 0.104    \\ 0.104 & 0.291  \end{array} \right)$ & $(-1.615, -1.425)$  &  &  &  &  &  \\

  \hline

    & 0.196 & $(-0.091, 0.912)$ & $\left( \begin{array}{cc} 0.548 & 0.074      \\   0.074 & 0.259    \end{array} \right)$ & $(-1.244, -1.396)$ &     &  &  &  &  \\
3SN & 0.228 & $(1.218, 0.158)$  & $\left( \begin{array}{cc} 0.807 & -0.181     \\ -0.181 & 0.467     \end{array} \right)$ & $(-1.203, 0.985 )$ & --- & $-2354.469$ & 17.733 & 58.758 & 23 \\
    & 0.576 & $(1.378, 0.516)$  & $\left( \begin{array}{cc} 0.455 & -0.028     \\ -0.028 & 0.218     \end{array} \right)$ & $(0.852, -0.683)$  &     &  &  &  &  \\

  \hline

\multirow{3}{*}{2T} & 0.277 & $(-0.231, 0.740)$ & $\left( \begin{array}{cc} 0.508 & -0.001   \\ -0.001 & 0.233  \end{array} \right)$ & --- & \multirow{3}{*}{11.195} & \multirow{3}{*}{$-2382.350$} & \multirow{3}{*}{49.497} & \multirow{3}{*}{23.59} & \multirow{3}{*}{12}  \\
                    & 0.723 & $(1.496, 0.404 )$ & $\left( \begin{array}{cc} 0.441 & 0.016    \\ 0.016 & 0.214   \end{array} \right)$ & --- &  &  &  &  &  \\

  \hline

   & 0.226 & $(-0.355, 0.758)$ & $\left( \begin{array}{cc} 0.458 & 0.016    \\ 0.016 &  0.223   \end{array} \right)$ & --- &  &  &  &  &  \\
3T & 0.156 & $(0.697, 0.372 )$ & $\left( \begin{array}{cc} 0.446 & -0.061   \\ -0.061 & 0.340   \end{array} \right)$ & --- & 16.391 & $-2351.602$ & 0. & 7.558 & 18 \\
   & 0.618 & $(1.588, 0.425)$  & $\left( \begin{array}{cc} 0.383 & -0.002   \\ -0.002 & 0.200   \end{array} \right)$ & --- &  &  &  &  &  \\

  \hline

\multirow{3}{*}{2ST} & 0.291 & $(-0.686, 0.830)$ & $\left( \begin{array}{cc} 0.752 & -0.056   \\ -0.056 & 0.247   \end{array} \right)$ & $(1.591, -0.569)$  & \multirow{3}{*}{12.089} & \multirow{3}{*}{$-2355.400$} & \multirow{3}{*}{3.596} & \multirow{3}{*}{0.} & \multirow{3}{*}{16} \\
                     & 0.709 & $(1.850, 0.531)$  & $\left( \begin{array}{cc} 0.552 & 0.069    \\ 0.069 & 0.240    \end{array} \right)$ & $(-1.361, -0.877)$ &  &  &  &  &  \\

  \hline

    & 0.194 & $(-0.230, 0.924)$ & $\left( \begin{array}{cc} 0.471 & 0.056    \\ 0.056 & 0.271    \end{array} \right)$ & $(-0.715, -1.462)$ &        &  &  &  &  \\
3ST & 0.196 & $(0.985, 0.231)$  & $\left( \begin{array}{cc} 0.628 & -0.167   \\ -0.167 & 0.411   \end{array} \right)$ & $(-1.025,  0.641)$ & 21.494 & $-2350.503$ & 9.802 & 50.827 & 24 \\
    & 0.610 & $(1.484, 0.479)$  & $\left( \begin{array}{cc} 0.397 & -0.010   \\ -0.010 & 0.208   \end{array} \right)$ & $(0.370, -0.335)$  &        &  &  &  &  \\

  \hline

\end{tabular}
\end{table*}

\begin{table}
\caption{The $\Delta_i$'s ($AIC$ and $BIC$), in increasing order, of the examined models for the BATSE data set.}
\label{tbl2}
\centering
\begin{tabular}{c c}
\hline
Model & $\Delta AIC$ \\
\hline
3T  &   0.    \\
2ST &   3.596 \\
3ST &   9.802 \\
3G  &  15.469 \\
3SN &  17.733 \\
2SN &  47.998 \\
2T  &  49.497 \\
2G  & 115.918 \\
\hline
\end{tabular}
\begin{tabular}{c c}
\hline
Model & $\Delta BIC$ \\
\hline
2ST &   0.    \\
3T  &   7.558 \\
3G  &  23.028 \\
2T  &  23.590 \\
2SN &  44.402 \\
3ST &  50.827 \\
3SN &  58.758 \\
2G  &  90.011 \\
\hline
\end{tabular}
\end{table}

\subsection{\it Fermi}
\label{sect3.2}

\begin{figure*}
\plotone{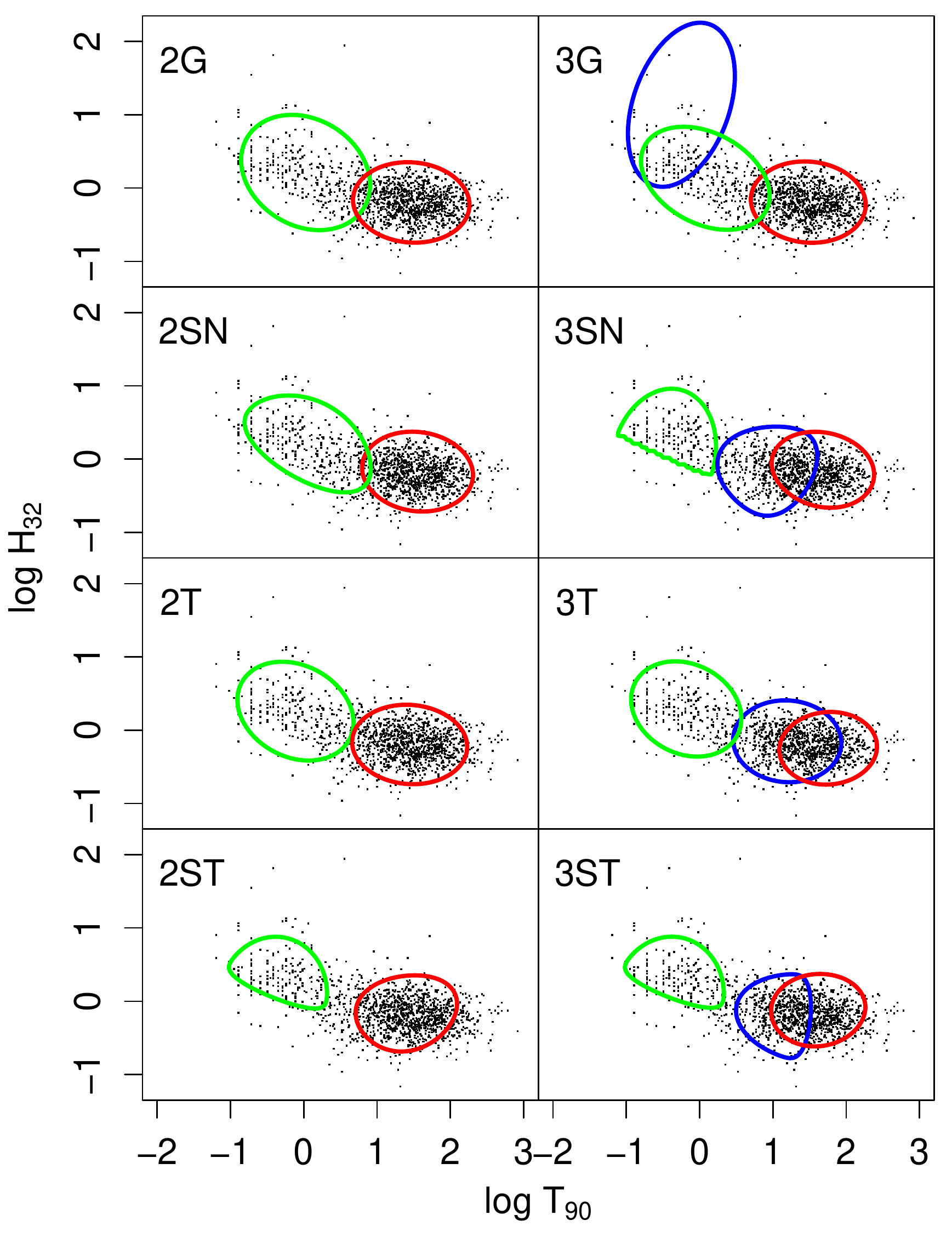}
\caption{Fits to the {\it Fermi} data. The contours depict the FWHM of each component, and $T_{90}$ is measured in seconds.}
\label{fig2}
\end{figure*}

In the same manner the 1376 {\it Fermi} GRBs were analyzed. The resulting parameters are gathered in Table~\ref{tbl3} and the fits are displayed in Fig.~\ref{fig2}. The two-component models are consistent with each other, just like most of the three-component models. A clear exception is the 3G fit, where the third component is not placed between the short and long GRBs, but drifts toward the harder part of the short GRBs' cluster. Setting different starting values for the fitting procedure or restricting the available range of parameters to force an outcome similar to the one obtained in case of BATSE data did not result in a quality fit---the $AIC$ and $BIC$ values were at best by a few hundreds greater than for the other models. The {\it Fermi} data set contains $\sim 15\%$ of short GRBs---this results in about half as many short GRBs as in the BATSE sample, which are also more sparse and spread out on the $T_{90}-H_{32}$ plane. Hence the weight of these points is high enough for the maximum loglikelihood procedure to take them into account when fitting.

\begin{table*}
\caption{Parameters of the fits to the {\it Fermi} data.}
\label{tbl3}
\centering
\begin{tabular}{c c c c c c c c c c}
\hline

Model & $A$ & $\bm{\mu}$ & $\bm{\Sigma}$ & $\bm{\lambda}$ & $\nu$ & $\mathcal{L}_{\rm max}$ & $\Delta AIC$ & $\Delta BIC$ & $p$ \\

  \hline

\multirow{3}{*}{2G} & 0.221 & $(0.025, 0.213 )$ & $\left( \begin{array}{cc} 0.558 & -0.107 \\ -0.107 & 0.446 \end{array} \right)$ & --- & \multirow{3}{*}{---} & \multirow{3}{*}{$-1525.740$} & \multirow{3}{*}{106.571} & \multirow{3}{*}{85.663} & \multirow{3}{*}{11} \\
                    & 0.779 & $(1.466, -0.196)$ & $\left( \begin{array}{cc} 0.455 & -0.019 \\ -0.019 & 0.217 \end{array} \right)$ & --- &  &  &  &  &  \\

  \hline

   & 0.009 & $(-0.248, 1.137)$ & $\left( \begin{array}{cc} 0.383 & 0.208  \\ 0.208 & 0.903   \end{array} \right)$ & --- &     &  &  &  &  \\
3G & 0.227 & $(0.078, 0.134 )$ & $\left( \begin{array}{cc} 0.556 & -0.151 \\ -0.151 & 0.356  \end{array} \right)$ & --- & --- & $-1478.543$ & 24.177 & 34.631 & 17 \\
   & 0.764 & $(1.483, -0.191)$ & $\left( \begin{array}{cc} 0.443 & -0.020 \\ -0.020 & 0.222  \end{array} \right)$ & --- &     &  &  &  &  \\

  \hline

\multirow{3}{*}{2SN} & 0.234 & $(-0.124, -0.162)$ & $\left( \begin{array}{cc} 0.614 & -0.049 \\ -0.049 & 0.579  \end{array} \right)$ & $(1.367, 2.359)$ & \multirow{3}{*}{---} & \multirow{3}{*}{$-1499.467$} & \multirow{3}{*}{62.024} & \multirow{3}{*}{60.956}  & \multirow{3}{*}{15} \\

                     & 0.766 & $(1.295, -0.211)$  & $\left( \begin{array}{cc} 0.483 & -0.016 \\ -0.016 & 0.215  \end{array} \right)$ & $(0.553, 0.161)$  &  &  &  &  &  \\

  \hline

    & 0.094 & $(-0.489, 0.038)$  & $\left( \begin{array}{cc} 0.367 & 0.070   \\ 0.070 & 0.615  \end{array} \right)$ & $(617.885, 1782.993)$ &     &  &  &  &  \\
3SN & 0.581 & $(1.378, -0.251)$  & $\left( \begin{array}{cc} 0.461 & -0.000  \\ -0.000 & 0.208 \end{array} \right)$ & $(0.812, 0.434)$      & --- & $-1470.025$ & 19.141 & 60.956 & 23 \\
    & 0.325 & $(1.457, -0.318)$  & $\left( \begin{array}{cc} 0.875 & -0.177  \\ -0.177 & 0.345 \end{array} \right)$ & $(-3.203, 1.058)$     &     &  &  &  &  \\

  \hline

\multirow{3}{*}{2T} & 0.187 & $(-0.114, 0.260)$ & $\left( \begin{array}{cc} 0.452 & -0.088 \\ -0.088 & 0.327 \end{array} \right)$ & --- & \multirow{3}{*}{12.118} & \multirow{3}{*}{$-1497.100$} & \multirow{3}{*}{49.29} & \multirow{3}{*}{28.383} & \multirow{3}{*}{12}  \\
                    & 0.813 & $(1.445, -0.195)$ & $\left( \begin{array}{cc} 0.445 -0.018   \\ -0.018 & 0.212 \end{array} \right)$ & --- &  &  &  &  &  \\

  \hline

   & 0.170 & $(-0.183, 0.290)$ & $\left( \begin{array}{cc} 0.408 & -0.071   \\ -0.071 & 0.304 \end{array} \right)$ & --- &  &  &  &  &  \\
3T & 0.494 & $(1.195, -0.151)$ & $\left( \begin{array}{cc} 0.392 & -0.012   \\ -0.012 & 0.225 \end{array} \right)$ & --- & 9.318 & $-1483.156$ & 33.403 & 43.857 & 18 \\
   & 0.336 & $(1.754, -0.247)$ & $\left( \begin{array}{cc} 0.320 & 0.012    \\ 0.012 & 0.177  \end{array} \right)$ & --- &  &  &  &  &  \\

  \hline

\multirow{3}{*}{2ST} & 0.121 & $(-0.423, 0.059)$ & $\left( \begin{array}{cc} 0.365 & 0.022  \\ 0.022 & 0.449  \end{array} \right)$ & $(2.552, 7.09)$  & \multirow{3}{*}{11.746} & \multirow{3}{*}{$-1468.454$} & \multirow{3}{*}{0.} & \multirow{3}{*}{0.} & \multirow{3}{*}{16} \\
                     & 0.879 & $(1.869, -0.312)$ & $\left( \begin{array}{cc} 0.699 & -0.090 \\ -0.090 & 0.239 \end{array} \right)$ & $(-1.854, 0.797)$ &  &  &  &  &  \\

  \hline

    & 0.124 & $(-0.414, 0.060)$ & $\left( \begin{array}{cc} 0.368 & 0.019  \\ 0.019 & 0.447  \end{array} \right)$ & $(2.355, 6.746 )$ &        &  &  &  &  \\
3ST & 0.390 & $(1.502, -0.239)$ & $\left( \begin{array}{cc} 0.671 & -0.074 \\ -0.074 & 0.247 \end{array} \right)$ & $(-7.253, 0.421)$ & 11.803 & $-1463.872$ & 6.836 & 48.651 & 24 \\
    & 0.486 & $(1.888, -0.298)$ & $\left( \begin{array}{cc} 0.390 & -0.047 \\ -0.047 & 0.216 \end{array} \right)$ & $(-0.765, 0.648)$ &        &  &  &  &  \\

  \hline

\end{tabular}
\end{table*}

\begin{figure}
\plotone{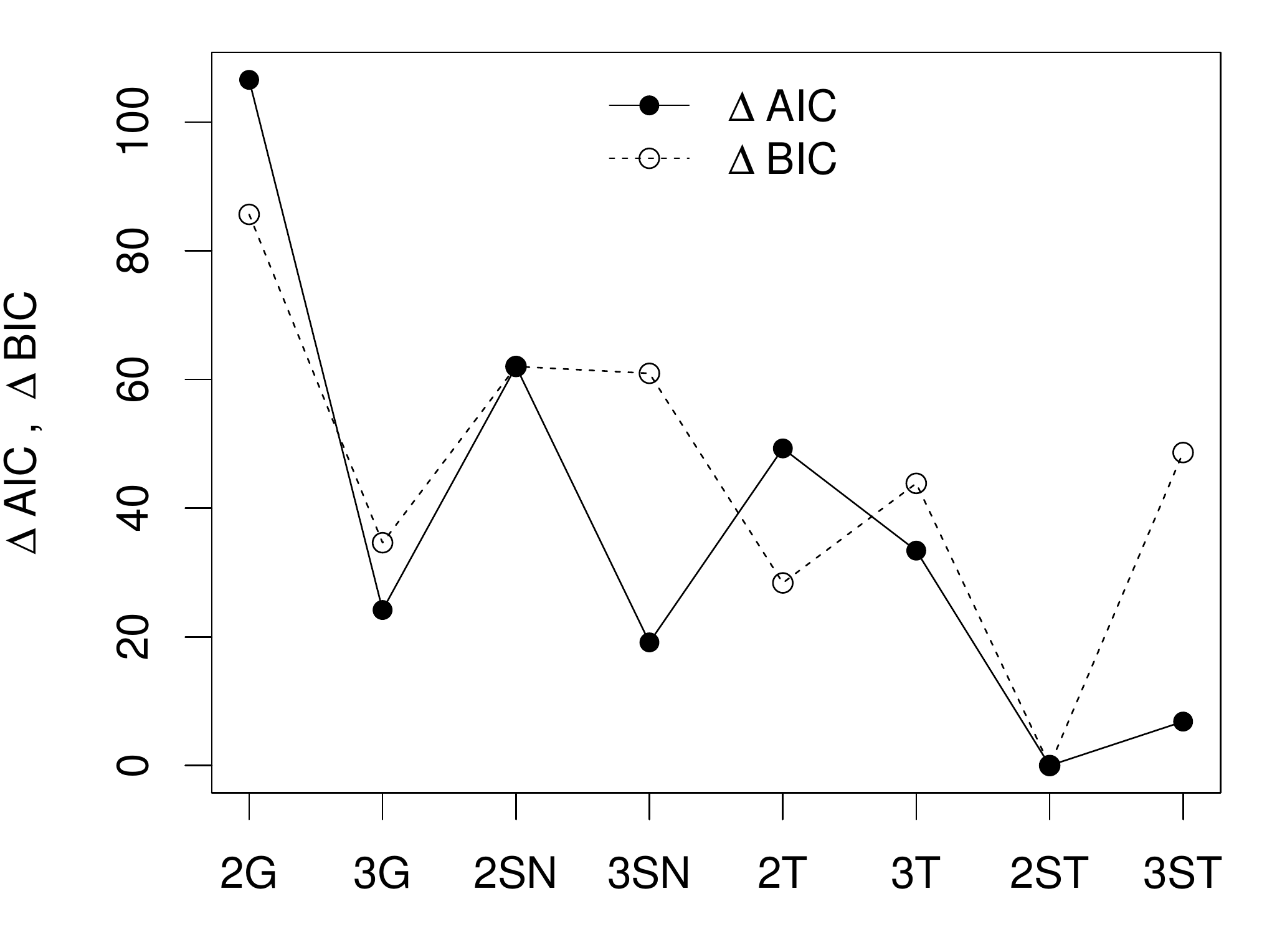}
\caption{Information criteria scores for the {\it Fermi} data.}
\label{fig4}
\end{figure}

Nonetheless, as follows from the $\Delta_i$ values from Fig.~\ref{fig4} and Table~\ref{tbl4}, the 2G model is again the worst among the examined, and the 3G is off the rate with $\Delta_i$ much higher than 10. Both $AIC$ and $BIC$ unanimously point at the 2$\mathcal{ST}$ as the best description of the data. In terms of $BIC$ no other model is even competitive, while there is moderate support for the 3$\mathcal{ST}$ in terms of $AIC$ ($\Delta_i = 6.84$). Overall, the 2$\mathcal{ST}$ is again the best description of the observed $(T_{90},H_{32})$ data points in the logarithmic plane. Note also that it has a comparable number of free components (16 vs. 17) to the previously broadly employed 3G, but does not invoke a new class of GRBs.

\begin{table}
\caption{The $\Delta_i$'s ($AIC$ and $BIC$), in increasing order, of the examined models for the {\it Fermi} data set.}
\label{tbl4}
\centering
\begin{tabular}{c c}
\hline
Model & $\Delta AIC$ \\
\hline
2ST &   0.    \\
3ST &   6.836 \\
3SN &  19.141 \\
3G  &  24.177 \\
3T  &  33.403 \\
2T  &  49.290 \\
2SN &  62.024 \\
2G  & 106.571 \\
\hline
\end{tabular}
\begin{tabular}{c c}
\hline
Model & $\Delta BIC$ \\
\hline
2ST &   0.    \\
2T  &  28.383 \\
3G  &  34.631 \\
3T  &  43.857 \\
3ST &  48.651 \\
3SN &  60.956 \\
2SN &  62.024 \\
2G  &  85.663 \\
\hline
\end{tabular}
\end{table}

\section{Discussion}
\label{sect4}

Up to date, all parametric analyses of the GRB population in the $T_{90}-H_{32}$ plane were conducted by means of the Gaussian mixture model. On this basis, the existence of a third class of GRBs, intermediate in durations and with soft spectra, has been claimed several times (see Sect.~\ref{sect1}). However, a similar claim based on the $\log T_{90}$ distribution alone was refuted by showing that the observed durations are better fitted by only two skewed components rather than three Gaussian ones \citep{tarnopolski15b,tarnopolski16a,tarnopolski16c}. The first to notice that the employment of Gaussian distributions is not fully justified were \citet{koen}, who wrote\footnote{However, \citet{mukh} noted that ''the distributions often seem bimodal with asymmetrical non-Gaussian shapes'', but failed to employ skewed distributions in modeling and proceeded considering ''the hypothesis that the sample consists of two or more distinct classes'' by assuming multinormal distributions.}: ''There is no guarantee that the components of a mixture correspond to physically distinct classes of objects. It is entirely possible that the distributions of class properties, such as $\log T_{90}$, are non-normal: in such a case, spurious classes would be identified due to the modelling of a non-normal distribution by normal components.'' Additionally, they showed in their fig.~14 two fits to {\it Swift} duration data that are very similar to each other, yet composed of entirely different components. Hence the association of a component of a statistical mixture to a physical class of objects and the inference of their properties is a dubious approach. \citet{zitouni} later suggested that the asymmetry in the duration distribution might come from a possible asymmetric distribution of the progenitor envelope mass. In this spirit, four bivariate statistical models (Sect.~\ref{sect2.3}) were tested herein: the Gaussian one, its skewed version (the $\mathcal{SN}$ distribution), the Student distribution, $\mathcal{T}$ (which, while being symmetric like the normal distribution, has a wider spread and a more slender shape), and its skewed version (the $\mathcal{ST}$ distribution). It was found (Sect.~\ref{sect3}) that despite, rather surprisingly, the mixture of $\mathcal{SN}$ distributions is not competitive with the Gaussian model (contrary to the univariate case; \citealt{tarnopolski16a}), the 2$\mathcal{ST}$ is the best description of the data among the examined possibilities. Particularly, it is a significant improvement of the fit compared to the 3G. The IC also indicate that the 3$\mathcal{ST}$ model is excessive. It should be emphasized that if the empirical distributions were not inherently skewed, this would be reflected in the fitting by obtaining $\bm{\lambda}\approx 0$ at least for some components of the mixtures, but this is not the case for neither the $\mathcal{SN}$, nor the $\mathcal{ST}$ models, regardless of the number of components employed (i.e., two or three). Therefore, the results imply that the existence of the presumed third GRB class, as a fundamentally distinct one from the short and long ones, is unlikely. On one hand, it concords with the possibility that this class may be in fact attributed, at least partially, to X-ray flashes (XRFs, \citealt{heise01,heise03,kippen,sakamoto05}) related to long GRBs \citep{sakamoto08}, and hence constitute the tail of the long GRBs group \citep{ripa14,ripa16}, especially in case of {\it Swift} GRBs (\citealt{veres}; but see also the discussion in \citealt{ripa16}). On the other hand, the presumed third class of GRBs, as observed by {\it RHESSI}, is not located at the soft tail of long GRBs, but between the short and long ones, hence is on average harder than XRFs \citep{ripa12}. In fact, their hardness ratios are comparable to those of short GRBs \citep{ripa16}. The smallest fraction of GRBs consistent with the definition of XRFs is in the BATSE catalog \citep{ripa16}. This shows that the intermediate class is indeed elusive, and its characteristics---in particular the location in the $T_{90}-H$ plane---are strongly detector-dependent, so any claims about its physical properties should be taken with caution.

The instrumental effects cannot be neglected in discussing the properties of GRB classes \citep{tarnopolski15b,ripa16}. {\it Swift} is more sensitive in soft bands compared to BATSE, hence it is more inclined towards detecting long GRBs and its low-luminosity tail---the putative intermediate class---than short ones. {\it BeppoSAX} is also more sensitive to long GRBs \citep{horvath09}, hence the lack of a distinct short GRB peak in the duration distribution. On the other hand, {\it Fermi} is more sensitive at very soft and very hard GRBs, yet a soft-intermediate tail of long GRBs is not visible in Fig.~\ref{fig2}---but the third component in the 3G model stretches from the short GRBs towards even higher hardness ratios. Except for this, in both BATSE and {\it Fermi} the third component is being located between the short and long groups, with typical hardness similar to long GRBs', especially in case of {\it Fermi} (Fig.~\ref{fig1} and \ref{fig2}). Different energy-detection intervals (e.g. $15-150\,{\rm keV}$ for {\it Swift}, and $8-1000\,{\rm keV}$ for {\it Fermi}) lead to contrasting group characteristics. {\it Suzaku}'s energy range ($50-5000\,{\rm keV}$) makes the resulting $T_{90}$ distribution similar to those of BATSE and {\it Fermi} rather than {\it Swift} \citep{ohmori}. Also, different energy sensitivities of the detectors give different estimates of $T_{90}$, and the flux limit for detection introduces a selection bias. Likewise, there are known observational and instrumental selection effects \citep{coward} related with the redshift distribution \citep{meszaros06} that affect the observed GRB samples. Finally, the spectrum of a GRB depends on the detector \citep{sakamoto11}, which in turn affects the calculated hardness ratio. Overall, the distinction between short and long GRBs---on observational, statistical and astrophysical grounds---is firm. The presumed third---intermediate in duration---class is putative, and there is no need to invoke it to describe the observed distributions of GRB properties. Lastly, a more flexible model, given by a mixture of copulas \citep{koen17}, would allow to separately model the marginals of a multivariate distribution, taking account of physical constraints on each variable independently.

Ideally, it is desirable to have the exact shape of the observed distributions derived from a physical theory, which has not been convincingly realized, though. However, as the 2G is better than 3G in the rest frame \citep{huja,tarnopolski16b,zhang16,kulkarni}, and competitive with skewed models \citep{tarnopolski16c} in the univariate case of $\log T_{90}$, it is hereby suggested that the redshift distribution of GRBs \citep{natarajan,meszaros06,li,meszaros11,le} is crucial in explaining the skewness of the observed quantities (Tarnopolski, in preparation). Some works \citep{mukh,roiger00,balastegui,chattopadhyay07,ripa12,chattopadhyay17,chattopadhyay18,acuner} have claimed to find three and more GRB groups in high dimensional parameter spaces. Such an approach must be undertaken with care, as {\it (i)} prinicipal component analyses usually pointed at three variables \citep{bagoly98,borgonovo06,bagoly09,horvath12,chattopadhyay17,acuner}, and {\it (ii)} higher dimensional spaces become more and more capacious, hence the identification of e.g. five clusters might be spurious.

\section{Summary}
\label{sect5}

\begin{enumerate}
\item Mixtures of two and three components of bivariate distributions: Gaussian, skew-normal ($\mathcal{SN}$), Student $t$ ($\mathcal{T}$), and skew-$t$ ($\mathcal{ST}$), were fitted to the $\log T_{90}-\log H_{32}$ data of {\it CGRO}/BATSE and {\it Fermi}/GBM.
\item Information criteria ($AIC$ and $BIC$) were used to establish that the 2$\mathcal{ST}$ model is significantly better in describing the data than any other among the considered ones. 
\item This is evidence for the non-existence of the elusive third, intermediate in durations and with soft spectra, class of GRBs.
\item The distributions of the GRBs' observed parameters are likely to be intrinsically skewed, possibly by the intrinsic skewness of the parameters governing the physical mechanism of a GRB.
\item It is suggested that the redshift distribution plays a crucial role in explaining the skewness of the parameters in the observer frame.
\end{enumerate}

\acknowledgments
The author is grateful to Peter Veres for providing the hardness ratios of {\it Fermi}/GBM GRBs, and wishes to thank Jakub \v R\'{\i}pa for comments on the manuscript. The final version of this manuscript was concluded during a visit at the Astronomical Institute of Charles University, Prague, Czech Republic. Support by the Polish National Science Center through an OPUS Grant No. 2017/25/B/ST9/01208 is acknowledged.

\software{\texttt{R} (\url{http://www.R-project.org/}), \texttt{mixsmsn} \citep[][\url{https://cran.r-project.org/web/packages/mixsmsn/index.html}]{prates}.}

\bibliography{bibliography}

\end{document}